\documentclass{llncs}

\usepackage{subfigure}
\usepackage{amsmath}
\usepackage{oz}
\usepackage{refcal}
\usepackage{url}
\usepackage{rcp}

\usepackage{xcolor}
\definecolor{dkgreen}{rgb}{0,0.6,0}

\usepackage{relsize}

\usepackage{time}
\usepackage{rcoms}
\usepackage{counters}
\usepackage{opsem}
\usepackage{multicol}
\usepackage{pdfsync}
\usepackage{wmm}
\usepackage{listings}

\newcommand{\ruledef}[3]{
	\begin{minipage}[b]{#1}
	\begin{equation}
		\label{rule:#2}
			#3
	\end{equation}
	\end{minipage}
}

\renewcommand{\refeqn}[1]{(\ref{eq:#1})}
\renewcommand{\refeqns}[2]{(\ref{eq:#1},\,\ref{eq:#2})}
\renewcommand{\WHbc}{\While b~\Do~c}

\renewcommand{\atomic}[1]{\seqT{#1}}
\def\atomicsep{~,~}

\renewcommand{\refrule}[1]{Rule~(\ref{rule:#1})}
\renewcommand{\reflaw}[1]{Law~(\ref{law:#1})}
\newcommand{\reflaws}[2]{Laws~(\ref{law:#1}) and~(\ref{law:#2})}

\newcommand{\refrulea}[1]{Rule~(\ref{rule:#1}a)}
\newcommand{\refruleb}[1]{Rule~(\ref{rule:#1}b)}

\newcommand{\addrShift}[2]{[#2,#1]}

\newcommand{\seqT}[1]{\langle #1 \rangle}

\newcommand{\srqmxv}[1]{\srq{\pidm}{x}{v}{#1}}
\newcommand{\srqmxvns}{\srqmxv{\pidns}}

\newcommand{\wseq}{W}

\newcommand{\pid}[1]{\textsc{#1}}
\newcommand{\pidn}{\pid{n}}
\newcommand{\pidm}{\pid{m}}
\newcommand{\pidns}{\pid{ns}}

\newcommand{\canReorder}[2]{#1 \ro #2}
\newcommand{\flush}{flush}
\newcommand{\PIDSet}{\textsc{PID}}

\newcommand{\srq}[4]{(#2 \mapsto #3)^{#1}_{#4}}
\newcommand{\srqnxv}[1]{\srq{\pidn}{x}{v}{#1}}
\newcommand{\srqnxvns}{\srqnxv{\pidns}}
\newcommand{\srqnxvn}{\srqnxv{\{\pidn\}}}
\newcommand{\loadgate}{\fence_L}
\newcommand{\storegate}{\fence_S}

\newcommand{\gvar}{\varphi}
\newcommand{\Read}[2]{\guard{#1 = #2}}
\newcommand{\Readx}[1]{\Read{x}{#1}}
\newcommand{\Readxv}{\Read{x}{v}}
\renewcommand{\eval}[2]{{#2}_{#1}}
\newcommand{\evalse}{\evals{e}}

\lstdefinelanguage{IMP}{
  	basicstyle={\normalfont},
  keywordstyle={\normalfont\bfseries},
	morekeywords={initially,locals,if,then,else,while,do,repeat,until,@,lcl},
	morekeywords={guard,fence,cfence,lwfence,CAS,lwfence},
	morekeywords={STR,LDR,eor,add},
	morekeywords={store,load,reg},
	morekeywords={thread,storage,globals},
	frame=none,
	mathescape=true,
	numbersep=0pt,
	comment=[l]{//},
  	commentstyle=\color{dkgreen},
    sensitive=false,
	numbers=none,
}

\newcommand{\code}[1]{$#1$}
\newcommand{\codesmall}[1]{\lstinline[basicstyle=\footnotesize\ttfamily]!#1!}

\newcommand{\lwfenced}[1]{\mathsf{lwf}(#1)}
\newcommand{\lwfencedn}{\lwfenced{\pidn}}

\renewcommand{\Update}[3]{{#1}_{[#2 \asgnsmall #3]}}
\newcommand{\Repl}[3]{{#1}_{[#2\backslash#3]}}

\renewcommand{\ttdef}{\mathbin{:\!:\!=}}

\renewcommand{\cbar}{\mathbin{~|~}}

\newcommand{\mathttbf}[1]{\mbox{\code{#1}}}
\newcommand{\mathttbfsmall}[1]{\mbox{\codesmall{#1}}}

\renewcommand{\If}{\mathrel{\keywordfont{if}}}
\renewcommand{\Then}{\mathrel{\keywordfont{then}}}
\renewcommand{\Else}{\mathrel{\keywordfont{else}}}
\renewcommand{\While}{\mathrel{\keywordfont{while}}}
\renewcommand{\Do}{{\mathrel{\keywordfont{do}}}}

\newcommand{\wpre}[2]{wp(#1, #2)}

\newcommand{\CAS}[3]{CAS(#1, #2, #3)}

\newcommand{\MT}[1]{$\mathtt{#1}$}

\newcommand{\reorderable}{\mathrel{\overset{\textsc{r}}{\Leftarrow}}}

\newcommand{\reordOf}[2]{#1 \leadsto #2}

\newcommand{\reordOfccp}{\reordOf{\cmdc}{\cmdc'}}

\newcommand{\nreorderable}{\not\!\reorderable}
\newcommand{\ro}{\reorderable}
\newcommand{\nro}{\nreorderable}

\renewcommand\notin{\not\,\!\in}

\newcommand{\ltiff}[1]{\quad \mbox{iff #1}}

\newcommand{\fwd}[2]{{#2}_{[#1]}}

\newcommand{\storage}[2]{(\keywordfont{stg}~ #1 @ #2)}
\newcommand{\thread}[2]{(\keywordfont{tid}_{#1} ~ #2)}

\newcommand{\lfence}{\keywordfont{lwfence}}

\newcommand{\befSkip}{}
\newcommand{\cbefSkip}{}

\newcommand{\defaultFont}[1]{#1}

\newcommand{\cmd}[1]{\defaultFont{#1}}
\newcommand{\cmdc}{\cmd{c}}
\newcommand{\cmdd}{\cmd{d}}

\newcommand{\ac}[1]{\defaultFont{#1}}
\newcommand{\aca}{\ac{\alpha}}
\newcommand{\acb}{\ac{\beta}}

\newcommand{\pr}[1]{\defaultFont{#1}}
\newcommand{\prp}{\pr{p}}
\newcommand{\cbef}{\mathbin{\mathtt{;}}}
\newcommand{\bef}{\,\centerdot\,}

\renewcommand{\Skip}{\mathbf{nil}}
\renewcommand{\asgn}{\mathbin{\mathtt{:\!=}}}

\def\keywordfont{\mathbf}

\renewcommand{\RegStoreWord}{{\mathttbf{MOV}}}
\renewcommand{\LoadWord}{{\mathttbf{LDR}}}
\renewcommand{\StoreWord}{{\mathttbf{STR}}}
\renewcommand{\RMWWord}{{\mathttbf{rmw}}}

\renewcommand{\RegStoreWord}{{\mathttbf{reg}}}
\renewcommand{\LoadWord}{{\mathttbf{load}}}
\renewcommand{\StoreWord}{{\mathttbf{store}}}
\renewcommand{\RMWWord}{{\mathttbf{rmw}}}

\renewcommand{\regstore}[2]{\mathtt{\RegStoreWord~#1 , #2}}
\renewcommand{\load}[2]{\mathtt{\LoadWord~#1 , #2}}
\renewcommand{\store}[2]{\mathtt{\StoreWord~#1 , #2}}
\renewcommand{\rmw}[3]{\mathtt{\RMWWord~#1,#2,#3}}
\renewcommand{\fence}{\keywordfont{fence}}
\renewcommand{\cfence}{\keywordfont{cfence}}
\newcommand{\guard}[1]{[#1]}
\newcommand{\guarde}{\guard{e}}
\newcommand{\guardb}{\guard{b}}

\newcommand{\guardlbl}[1]{[#1]}
\newcommand{\guardlble}{\guardlbl{e}}

\newcommand{\fencelbl}{\fence}

\newcommand{\asgnlbl}{\mathrel{\,\mathtt{:=}\,}}

\newcommand{\storegatelbl}{\mathttbfsmall{fence}_S}
\newcommand{\loadlbl}[2]{\mathttbfsmall{load}(#1, #2)}

\newcommand{\localsWord}{\keywordfont{lcl}}
\newcommand{\localsnp}[2]{\localsWord~#1 @ #2} 
\renewcommand{\locals}[2]{(\localsnp{#1}{#2})}
\newcommand{\globals}[2]{(\keywordfont{glb}~#1 @ #2)}

\renewcommand{\prefix}[2]{#1 \cbef #2}
\renewcommand{\prefixa}[1]{\prefix{\aca}{#1}}
\renewcommand{\prefixac}{(\prefixa{\cmdc})}
\renewcommand{\prefixacp}{(\prefixa{\cmdc'})}

\usepackage{anyfontsize}

\title{A wide-spectrum language for verification of programs on weak memory models}
\author{Robert J. Colvin and Graeme Smith}
\institute{School of Information Technology and Electrical Engineering \\ University of Queensland}

\begin{document}
\maketitle

\lstset{language=IMP}

\begin{abstract}

Modern processors deploy a variety of weak memory models, which for efficiency reasons may (appear to) execute instructions in an order different to
that specified by the program text.  The consequences of instruction reordering can be complex and subtle, and can impact on ensuring
correctness.
Previous work on the semantics of weak memory models has focussed on the behaviour of assembler-level programs.  In this
paper we utilise that work to extract some general principles underlying instruction reordering, and apply those principles
to a wide-spectrum language
encompassing abstract data types as well as low-level assembler code
.  The goal is
to support reasoning about implementations of data structures for modern processors with respect to an abstract specification.  

Specifically, we define an operational semantics, from which we derive some properties of program refinement, and encode the semantics in the rewriting
engine Maude as a model-checking tool.  The tool is used to validate the semantics against the behaviour of a set of litmus tests (small
assembler programs) run on hardware, and also to model check implementations of data structures from the literature against their
abstract specifications.

\end{abstract}

\section{Introduction} 

Modern processor architectures provide a challenge for developing efficient and correct software.  Performance can
be improved by parallelising computation to utilise multiple cores, but communication between threads is notoriously error-prone.
Weak memory models go further and improve overall system efficiency through sophisticated techniques for batching reads and writes to the
same variables and to and from the same processors.  However, code that is run on such memory models is not guaranteed to take effect in
the order specified in the program code, creating unexpected behaviours for those who are not forewarned \cite{AdveBoehm2010}.  For instance, the
instructions $x \asgn 1 \cbef y \asgn 1$ may be reordered to $y \asgn 1 \cbef x \asgn 1$.
Architectures typically provide {\em memory barrier\/}/{\em fence\/} instructions which can enforce local ordering -- so that $x \asgn 1 \cbef \fence
\cbef y \asgn 1$ can not be reordered -- 
but reduce performance improvements (and so should not be overused).  

Previous work on formalising weak memory models has resulted in abstract formalisations which were developed incrementally through
communication with processor vendors and rigorous testing on real machines \cite{UnderstandingPOWER,HerdingCats,ModellingARMv8}.  A large
collection of ``litmus tests'' have been developed \cite{LitmusTests,Mador-Haim2010} which demonstrate the sometimes confusing behaviour of hardware.  We utilise this
existing work to provide a wide-spectrum programming language and semantics that runs on the same relaxed principles that apply to assembler
instructions.  When these principles are specialised to the assembler of ARM and POWER processors our semantics gives behaviour consistent with existing litmus tests.
Our language and semantics, therefore, connect instruction reordering to higher-level notions of correctness. This enables verification of low-level code targeting specific processors against abstract specifications.

\OMIT{
Our theory is encoded in Maude \cite{Maude}, which provides an efficient rewriting engine.  
The operational semantics generates a trace consisting of a sequence of actions which may be loads, stores, fences, etc.
The behaviour of an architecture is specified as the relationship
between individual types of actions, in addition to the behaviour of the storage system.
}

We begin in \refsect{overview} with the basis of an operational semantics that allows reordering of instructions according to pair-wise
relationships between instructions.  
In \refsect{semantics} we
describe the semantics in more detail, focussing on its instantiation for the widely used ARM and POWER processors.
In \refsect{tool} we give a summary of the encoding of the semantics in Maude and its application to model-checking concurrent data structures.
We discuss related work in \refsect{related-work} before concluding in \refsect{conclusion}.

\OMIT{
In \refsect{models}
we show the instantiations of the thread-local definitions to three well-known
weak memory models, TSO \cite{x86-TSO}, ARM \cite{ModellingARMv8} and POWER \cite{UnderstandingPOWER}.  
We then consider the implications of
weak memory models on more complex algorithms in \refsect{higher-level-code}: we verify a simple lock
\cite[Sect. 7.3]{HerlihyShavit2011}, the Treiber lock-free stack \cite{Treiber86} running on ARM and POWER, and find (and fix) a bug in an implementation of the
Chase-Lev work-stealing deque (double-ended queue) \cite{ChaseLev05} developed specifically for ARM \cite{LeWorkStealingPPoPP13}.  
}

\section{Instruction reordering in weak memory models}
\label{overview}

\subsection{Thread-local reorderings}

It is typically 
assumed processes are executed in a fixed sequential order (as given by sequential composition -- the ``program order'').  
However program order may be inefficient, 
e.g., when retrieving the value of a variable from main memory after setting its value, as in \MT{x \asgn 1 \cbef r \asgn x}, and hence
weak memory models sometimes allow execution out of program order to improve overall system efficiency.
While many reorderings can seem surprising, there are basic principles at play which limit the number of possible permutations, the key
being that the new ordering of instructions preserves the original sequential intention.

A classic example of weak memory models producing unexpected behaviour is the ``store buffer'' pattern below \cite{LitmusTests}.
Assume that all variables are initially 0, and that thread-local variables (registers) are named $r, r_1, r_2$,
etc., and that $x$ and $y$ are shared variables.
\begin{equation}
	(x \asgn 1 \cbef r_1 \asgn y)
	\pl 
	(y \asgn 1 \cbef r_2 \asgn x)
\end{equation}
It is possible to reach a final state in which $r_1 = r_2 = 0$ in several weak memory models:  the two assignments in each
process are independent (they reference different variables), and hence can be reordered.  From a sequential semantics perspective,
reordering the assignments in process 1, for example, preserves the final values
for $r_1$ and $x$.

Assume that $c$ and $c'$ are programs represented as sequences of atomic actions
$\aca \cbef \acb \cbef \ldots$, as in a sequence of instructions in a processor or more abstractly a semantic trace.  
Program $c$ may be reordered to $c'$, written $\reordOfccp$, if the following holds:
\begin{enumerate}
\item
$\cmdc'$ is a permutation of the actions of $\cmdc$, possibly with some modifications due to \emph{forwarding} (see below).

\item
$\cmdc'$ preserves the sequential semantics of $\cmdc$.  For example, in a weakest preconditions semantics \cite{GuardedCommands},
$(\all S @ \wpre{\cmdc}{S} \imp \wpre{\cmdc'}{S})$.

\item
$\cmdc'$ preserves \emph{coherence-per-location} with respect to $\cmdc$
(cf. \T{po-loc} in \cite{HerdingCats}).
This means that
the order of updates and accesses of each shared variable, considered individually, is maintained.
\end{enumerate}
We formalise these constraints below.
The key challenge for reasoning about programs executed on a weak memory model is that
the behaviour of $\cmdc \pl \cmdd$ is in general quite different to the behaviour of $\cmdc' \pl \cmdd$, even if $\reordOfccp$.

\subsection{Reordering and forwarding instructions}
\label{overview-reordering}
\OMIT{
\begin{equation}
\begin{split}
	~&
	\reordOf{
	(x \asgn e \cbef y \asgn f)
	}{
	(y \asgn f \cbef x \asgn e )
	}
	\qquad
	if
	\\
		~&
	\parbox{\textwidth-10mm}{
		1) $x$, $y$ are distinct; 
		2) $e$, $f$ are load-distinct; 
		3) $y \nfi e$; and 
		4) $x \nfi f$
	}
\end{split}
\end{equation}
The constraint disallows the reordering of
$(r \asgn 1 \cbef x \asgn r)$ to $(x \asgn r \cbef r \asgn 1)$, which violates the ``$x \nfi f$'' ($r \nfi r$) condition above.
}

We write $\alpha \ro \beta$ if instruction $\beta$ may be reordered before instruction $\alpha$.
It is relatively straightforward to define when two assignment instructions (encompassing stores, loads, and register operations at the assembler level)
may be reordered. 
Below let $x \nfi f$ mean that $x$ does not appear free in the expression $f$,
and say expressions $e$ and $f$ are \emph{load-distinct} if they do not reference any common shared variables.
\begin{equation}
\begin{split}
\label{eq:reordering-principle}
	~&
	x \asgn e \ro y \asgn f
	\quad
	if
	\quad
	\parbox{0.6\textwidth}{
		1) $x$, $y$ are distinct; 
		~~2) $x \nfi f$; 
		~~3) $y \nfi e$; 
		and \\
		4) $e$, $f$ are load-distinct; 
	}
\end{split}
\end{equation}
Note that in general $\ro$ is not reflexive:  in TSO processors a load may be reordered before a store, but not vice versa \cite{x86-TSO}.

Provisos 1), 2) and 3) ensure executing the two assignments in either order results in the same final values for $x$ and $y$, and proviso 4)
maintains order on accesses of the shared state.
If two updates do not refer to any common variables they may be reordered.  The provisos allow some reordering when they share common
variables.
Proviso 1) eliminates reorderings such as
$
	\reordOf{
		(x \asgn 1 \cbef x \asgn 2)
	}{
		(x \asgn 2 \cbef x \asgn 1)
	}
$ 
which would violate the sequential semantics (the final value of $x$).
Proviso 2) eliminates reorderings such as
$
	\reordOf{
		(x \asgn 1 \cbef r \asgn x)
	}{
		(r \asgn x \cbef x \asgn 1)
	}
$
which again would violate the sequential semantics (the final value of $r$).
Proviso 3) eliminates reorderings such as
$
	\reordOf{
		(r \asgn y \cbef y \asgn 1)
	}{
		(y \asgn 1 \cbef r \asgn y)
	}
$
which again would violate the sequential semantics (the final value of $r$).
Proviso 4), requiring the update expressions to be load-distinct, preserves coherence-per-location,
eliminating reorderings such as
$
	\reordOf{
		(r_1 \asgn x \cbef r_2 \asgn x)
	}{
		(r_2 \asgn x \cbef r_1 \asgn x)
	}
$, where $r_2$ may receive an earlier value of $x$ than $r_1$ in an environment which modifies $x$.

In practice, proviso 2) may be circumvented by \emph{forwarding}%
\footnote{We adopt the term ``forwarding" from ARM and POWER \cite{HerdingCats}. The equivalent effect is referred to as \emph{bypassing} on TSO \cite{x86-TSO}.}%
.  This refers to taking into account the effect of the earlier update on the expression of the latter.
We write $\fwd{\alpha}{\beta}$ to represent the effect of forwarding the (assignment) instruction $\alpha$ to the instruction $\beta$.
For assignments we define
\begin{equation}
\label{eq:assignment-forwarding}
	\fwd{x \asgnsmall e}{(y \asgn f)}
	~~=~~
	y \asgn (\Repl{f}{x}{e})
	\quad
	if
	\quad
	\mbox{$e$ does not refer to global variables}
\end{equation}
where
the term $\Repl{f}{x}{e}$ stands for the syntactic replacement in expression $f$ of references to $x$ with $e$.
The proviso of \refeqn{assignment-forwarding} prevents additional loads of globals being introduced by forwarding.

\OMIT{
The final constraint eliminates reorderings such as
$
	\reordOf{
		(r_1 \asgn x \cbef r_2 \asgn r_1)
	}{
		(r_2 \asgn x \cbef r_1 \asgn x)
	}
$.
Note that the assignment to $r_2$ has been affected by forwarding.  Here a second load of $x$ has been introduced, possibly giving
different final values for $r_1$ and $r_2$.  If instead $x$ was a local register, there would be no harm in the reordering (and forwarding)
since the environment may not modify the register.
The proviso does allow reorderings with forwarding such as
$
	\reordOf{(r \asgn 1 \cbef x \asgn r)}{(x \asgn 1 \cbef r \asgn 1)}
$.
}

We
specify the reordering and forwarding relationships with other instructions such as branches and fences in
\refsect{reordering-forwarding}.

\subsection{General operational rules for reordering}
\label{overview-rule}

The key operational principle allowing reordering is given by the following transition rules
for a program $(\aca \cbef \cmdc)$, i.e., a program with initial instruction $\aca$.
\begin{equation}
\label{rule:reorder-rule}
	\prefixac \tra{\aca} \cmdc
	~~~(a)
	\qquad
	\qquad
	\Rule{
		\cmdc \tra{\acb} \cmdc'
		\quad
		\aca \ro \fwd{\aca}{\acb}
	}{
		\prefixac \ttra{\fwd{\aca}{\acb}} \prefixacp
	}
	~~~(b)
\end{equation}
\refrulea{reorder-rule}
is the straightforward promotion of the first instruction into a step in a trace, similar to the basic prefixing rules of CCS
\cite{CCS} and CSP \cite{CSP}.
\refruleb{reorder-rule},
however, states that, unique to weak memory models, an instruction of $\cmdc$, say $\acb$, can happen before $\aca$, provided that
$\fwd{\aca}{\acb}$ can be reordered before $\aca$
according to the rules of the architecture.
Note that we forward the effect of $\aca$ to $\acb$ before deciding if the reordering is possible.

Applying \refruleb{reorder-rule} then \refrulea{reorder-rule} gives the following reordered behaviour of two assignments.
\begin{equation}
	(r \asgn 1 \cbef x \asgn r \cbef \Skip)
	\ttra{x \asgnsmall 1}
	(r \asgn 1 \cbef \Skip)
	\ttra{r \asgnsmall 1}
	\Skip
\end{equation}
We use the command $\Skip$ to denote termination.
The first transition above is possible because we calculate the effect of $r \asgn 1$ on the update of $x$ before executing that update, i.e.,
$\fwd{r \asgnsmall 1}{x \asgn r} = x \asgn 1$.
	
The definitions of instruction reordering, $\aca \ro \acb$, and instruction forwarding, $\fwd{\aca}{\acb}$
are architecture-specific, and are the only definitions required to specify an architecture's instruction ordering.%
\footnote{Different architectures may have different
storage subsystems, however, and these need to be separately defined (see \refsect{storage-subsystem-semantics}).}
The instantiations for {\em sequentially consistent\/} processors (i.e., those which do not have a weak memory model) are trivial: $\aca \nro
\acb$ for all $\aca, \acb$, and there is no forwarding.  Since reordering is not possible \refruleb{reorder-rule} never
applies and hence the standard prefixing semantics is maintained.
TSO is relatively straightforward: loads may be reordered before stores (provided they reference different shared variables).
In this paper we focus on the more complex ARM and POWER memory models. These memory models are very similar, the notable difference being the inclusion of the \emph{lightweight fence}
instruction in POWER. Due to space limitations, we omit lightweight fences in this paper but a full definition which has been validated against litmus tests can be found in \refappendix{power}.

\subsection{Reasoning about reorderings}

The operational rules allow a standard trace model of correctness to be adopted, that is, we say programs $\cmdc$ 
\emph{refines to} program $\cmdd$, written $\cmdc \refsto \cmdd$, iff every trace of $\cmdd$ is a trace of $\cmdc$.  
Let the program $\aca \bef \cmdc$ have the standard semantics of prefixing, that is, the action $\aca$ always occurs before any action in
$\cmdc$ (\refrulea{reorder-rule}).  Then we can derive the following laws that show the interplay of reordering
and true prefixing.
\begin{eqnarray}
	\aca \cbef \cmdc
	&\refsto&
	\aca \bef \cmdc
\label{law:keep-order}
	\\
	\aca \cbef (\acb \bef \cmdc)
	&\refsto&
	\fwd{\aca}{\acb} \bef (\aca \cbef \cmdc)
	\qquad
	\mbox{if $\aca \ro \fwd{\aca}{\acb}$}
\label{law:swap-order}
\end{eqnarray}
Note that in \reflaw{swap-order} $\aca$ may be further reordered with instructions in $\cmdc$.
A typical interleaving law is the following.
\begin{equation}
	(\aca \bef \cmdc) \pl \cmdd
	\refsto
	\aca \bef (\cmdc \pl \cmdd)
\label{law:fix-interleaving}
\end{equation}
We may use these laws to show how the
``surprise'' behaviour of the store buffer pattern above arises.%
\footnote{
To focus on instruction reorderings we leave local variable declarations and process ids implicit,
and assume a multi-copy atomic storage system (see \refsect{storage-subsystem-semantics}).
}
In derivations such as the following, to save space, we abbreviate a thread $\prefixa{\Skip}$ or $\aca \bef \Skip$ to $\aca$, that is, we omit the trailing $\Skip$.
\begin{derivation}
	\step{
		(x \asgn 1 \cbef r_1 \asgn y \cbefSkip)
		\pl 
		(y \asgn 1 \cbef r_2 \asgn x \cbefSkip)
	}
	\trans{\refsto}{
		From \reflaw{swap-order} (twice), since $x \asgn 1 \ro r_1 \asgn y$ from (\ref{eq:reordering-principle}).
	}
	\step{
		(r_1 \asgn y \bef x \asgn 1 \befSkip)
		\pl 
		(r_2 \asgn x \bef y \asgn 1 \befSkip)
	}
	\trans{\refsto}{
		\reflaw{fix-interleaving} (four times) and commutativity of $\pl$.
	}
	\step{
		r_1 \asgn y \bef r_2 \asgn x \bef x \asgn 1 \bef y \asgn 1 \befSkip
	}
\end{derivation}
If initially $x = y = 0$, a standard sequential semantics shows that $r_1 = r_2 = 0$ is a possible final state in this behaviour.

\section{Semantics}
\label{semantics}

\subsection{Formal language}

\newcommand{\sys}[1]{\defaultFont{#1}}
\newcommand{\syss}{\sys{s}}

The elements of our wide-spectrum language are actions (instructions) $\alpha$, commands (programs) $c$, processes (local state and a command) $p$, and the top level system $s$,
encompassing a shared state and all processes.
Below $x$ is a variable (shared or local) and $e$ an expression.
\begin{equation}
\begin{aligned}
	\aca &\ttdef 
		x \asgn e
		\cbar
		\guarde
		\cbar
		\fence
		\cbar
		\cfence
		\cbar
		\aca^*
	\\
	\cmdc &\ttdef
		\Skip 
		\cbar
		\aca \cbef \cmdc
		\cbar
		\cmdc_1 \choice \cmdc_2
		\cbar
		\WHbc
	\\
	\prp &\ttdef
		\locals{\sigma}{\cmdc}
		\cbar
		\thread{\pidn}{\prp}
		\cbar
		\prp_1 \pl \prp_2
	\\
	\syss &\ttdef
		\globals{\sigma}{\prp}
		\cbar
		\storage{\wseq}{\prp}
\end{aligned}
\end{equation}
An action may be an update $x \asgn e$, a guard $\guarde$, a (full) fence,
a control fence (see \refsect{reordering-forwarding}),
or a finite sequence of actions, $\aca^*$, executed atomically.  Throughout the paper we denote an empty sequence by
$\eseq$, and construct a non-empty sequence as $\seqT{\alpha_1 \atomicsep \alpha_2 \ldots}$.

A command may be the empty command $\Skip$, which is already terminated, a command prefixed by some action $\aca$, a choice between
two commands, or an iteration (for brevity we consider only one type of iteration, the while loop).
\OMIT{
Sequential composition of commands, as opposed to action prefixing, can be defined by induction.
\begin{eqnarray*}
	\Skip \scomp \cmdc = \cmdc
	\qquad
	\qquad
	&&
	\qquad
	\qquad
	(\aca \cbef \cmdc_1) \scomp \cmdc_2 = \aca \cbef (\cmdc_1 \scomp \cmdc_2)
	\\
	(\cmdc_1 \choice \cmdc_2) \scomp \cmdc_3 &=& (\cmdc_1 \scomp \cmdc_3) \choice (\cmdc_2 \scomp \cmdc_3)
\end{eqnarray*}
}
Conditionals are modelled using guards and choice.
\begin{equation}
\label{eq:defn-if}
	\IFbc \sdef (\guard{b} \cbef \cmdc_1) \choice (\guard{\neg b} \cbef \cmdc_2)
\end{equation}

A well-formed process is structured as a process id $\pidn \in \PIDSet$ encompassing a (possibly empty) local state $\sigma$ and command $\cmdc$, i.e., a term
$\thread{\pidn}{\localsnp{\sigma}{\cmdc}}$.
We assume that all local variables referenced in $\cmdc$ are contained in the domain of $\sigma$.

A system is structured as the parallel composition of processes within the global storage system, 
which may be either a typical global
state, $\sigma$, that maps all global variables to their values (modelling the storage systems of TSO, the most recent version of ARM, and abstract specifications),
\OMIT{
\begin{equation}
\label{eq:system-structure-sigma}
	\globals{\sigma}{
		\thread{1}{\localsnp{\sigma_1}{\cmdc_1}} 
		\pl
		\thread{2}{\localsnp{\sigma_2}{\cmdc_2}}
		\pl
		\ldots
	}
\end{equation}
}%
or a storage system, $\wseq$, formed from a list of ``writes'' to the global variables
(modelling the storage systems of older versions of ARM and POWER). 
Hence a system is in one of the two following forms.
\begin{equation}
\begin{aligned}
\label{eq:system-structure}
	\globals{\sigma}{
		\thread{1}{\localsnp{\sigma_1}{\cmdc_1}} 
		\pl
		\thread{2}{\localsnp{\sigma_2}{\cmdc_2}}
		\pl
		\ldots
	}
	\\
	\storage{\wseq}{
		\thread{1}{\localsnp{\sigma_1}{\cmdc_1}} 
		\pl
		\thread{2}{\localsnp{\sigma_2}{\cmdc_2}}
		\pl
		\ldots
	}
\end{aligned}
\end{equation}

\subsection{Operational semantics}

The meaning of our language is formalised using an operational semantics,
summarised in \reffig{fig:semantics-main}.
Given a program $\cmdc$ the operational semantics generates a \emph{trace}, \ie, a possibly infinite sequence of steps
$\cmdc_0 \tra{\aca_1} \cmdc_1 \tra{\aca_2} \ldots$ where the labels in the trace are actions, or  
a special label
$\tau$ representing a silent or internal step that has no observable effect.  

\begin{figure}[t]

\ruledef{80mm}
{reorder-rule-2}{
	\prefixac \tra{\aca} \cmdc
	~~~(a)
	\qquad
	\Rule{
		\cmdc \tra{\acb} \cmdc'
		\quad
		\aca \ro \fwd{\aca}{\acb}
	}{
		\prefixac \ttra{\fwd{\aca}{\acb}} \prefixacp
	}
	~~~(b)
}
\ruledef{30mm}
{nondet}{
	\begin{array}{lcl}
	\cmdc \choice \cmdd &\tra{\tau}& \cmdc
	\\
	\cmdc \choice \cmdd &\tra\tau{}& \cmdd
	\end{array}
}

\vskip 2.5mm

\ruledef{100mm}
{loop-unfold-rule}{
	\WHbc \tra{\tau} (\guard{b} \cbef \cmdc \scomp \,\, \WHbc) \choice (\guard{\neg b} \cbef \Skip)
}

\ruledef{60mm}
{locals-reg}{
	\Rule{
		\cmdc \ttra{r \asgnlbl v} \cmdc'
	}{
		\locals{\sigma}{\cmdc}
		\tra{\tau}
		\locals{\Update{\sigma}{r}{v}}{\cmdc'}
	}
}
\ruledef{58mm}
{locals-store}{
	\Rule{
		\cmdc \ttra{x \asgnlbl r} \cmdc'
		\quad
		\sigma(r) = v
	}{
		\locals{\sigma}{\cmdc}
		\ttra{x \asgnlbl v}
		\locals{\sigma}{\cmdc'}
	}
}

\ruledef{60mm}
{locals-load}{
	\Rule{
		\cmdc \ttra{r \asgnlbl x} \cmdc'
	}{
		\locals{\sigma}{\cmdc}
		\ttra{\Readxv}
		\locals{\Update{\sigma}{r}{v}}{\cmdc'}
	}
}
\ruledef{58mm}
{locals-guard}{
	\Rule{
		\cmdc \ttra{\guardlble} \cmdc'
	}{
		\locals{\sigma}{\cmdc}
		\ttra{\guardlbl{e_{\sigma}}}
		\locals{\sigma}{\cmdc'}
	}
}

\ruledef{44mm}
{threads}{
	\Rule{
		p \tra{\aca} p'
	}{
		\thread{\pidn}{p}
		\ttra{\pidn:\aca}
		\thread{\pidn}{p'}
	}
}
\ruledef{75mm}
{pl}{
	\Rule{
		\prp_1 \tra{\aca} \prp_1'
	}{
		\prp_1 \pl \prp_2 \tra{\aca} \prp_1' \pl \prp_2
	}
	\quad
	\Rule{
		\prp_2 \tra{\aca} \prp_2'
	}{
		\prp_1 \pl \prp_2 \tra{\aca} \prp_1 \pl \prp_2'
	}
}

\ruledef{65mm}
{globals-store}{
	\Rule{
		p \ttra{\pidn: x \asgnlbl e} p'
	}{
		\globals{\sigma}{p}
		\tra{\tau}
		\globals{\Update{\sigma}{x}{\evalse}}{p'}
	}
}
\ruledef{52mm}
{globals-guard}{
	\Rule{
		p \ttra{\pidn: \guardlble} p'
		\quad
		\evalse  \equiv true
	}{
		\globals{\sigma}{p}
		\tra{\tau}
		\globals{\sigma}{p'}
	}
}

\caption{Semantics of the language}
\label{fig:semantics-main}
\end{figure}


The terminated command $\Skip$ has no behaviour; a trace that ends with this command is assumed to have completed.
The effect of instruction prefixing in \refrule{reorder-rule-2} is discussed in \refsect{overview-rule}.
Note that actions become part of the trace.  We describe an instantiation for reordering and forwarding corresponding to the semantics
of ARM and POWER in \refsect{reordering-forwarding}.

A nondeterministic choice (the \emph{internal choice} of CSP \cite{CSP}) can choose either branch, as given by \refrule{nondet}.
The semantics of loops is given by unfolding, e.g., \refrule{loop-unfold-rule} for a `while' loop.
Note that {\em speculative execution\/}, i.e., early execution of instructions which occur after a branch point \cite{PrimerMemoryConsistency11}, is theoretically unbounded, and loads from inside later iterations of the loop could occur in earlier
iterations.  


For ease of presentation in defining the semantics for local states, we give rules for specific forms of actions, i.e.,
assuming that $r$ is a local variable in the domain of $\sigma$, and that $x$ is a global (not
in the domain of $\sigma$).
The more general version
can be straightforwardly constructed from the principles below.  

\refrule{locals-reg}
states that an action updating variable $r$ to value $v$ results in a change to the local state (denoted $\Update{\sigma}{r}{v}$).  Since this is a purely local operation there
is no interaction with the storage subsystem and hence the transition is promoted as a silent step $\tau$.
\refrule{locals-store} states that a \emph{store} of the value in variable $r$ to global $x$ is promoted as an instruction $x \asgn v$
where $v$ is the local value for $r$.
\refrule{locals-load} covers the case of a 
\emph{load} of $x$ into $r$.  The value of $x$ is not known locally.  
The promoted label is a
guard requiring that
the value read for $x$ is $v$.  This transition is possible for any value of $v$, but the correct value will be resolved when the label is
promoted to the storage level.
\refrule{locals-guard} states that a guard is partially evaluated with respect to the local state before it is promoted to the global level.  The notation $\evalse$ replaces $x$ with $v$ in $e$ for all $(x \mapsto v) \in \sigma$.

\refrule{threads} simply tags the process id to an instruction, to assist in the interaction with the storage system, and otherwise has no effect. Instructions of concurrent processes are interleaved in the usual way as described by \refrule{pl}.  

Other straightforward rules which we have omitted above include the promotion of
fences through a local state, and that atomic sequences of actions are handled inductively by the above rules. 

\subsubsection{Multi-copy atomic storage subsystem.}
\label{storage-subsystem-semantics}

Traditionally, changes to shared variables occur on a shared global state, and when written to the global state are seen instantaneously by all processes in the system.  This is referred to as {\em multi-copy atomicity\/} and is a feature of TSO and the most recent version of ARM \cite{ARMv8.4}.
Older versions of ARM and
POWER, however, lack such multi-copy atomicity and require a more complex semantics.  We give the simpler case (covered in \reffig{fig:semantics-main}) 
first.%
\footnote{
In this straightforward
model of shared state there is no global effect of fences, and we omit the straightforward promotion rule.  
}

Recall that at the global level the process id $\pidn$ has been tagged to the actions by \refrule{threads}.  
\refrule{globals-store} covers a store of some expression $e$ to $x$.  
Since all local variable references have been replaced by their values at the process level due to Rules~(\ref{rule:locals-reg})-(\ref{rule:locals-guard}),
expression $e$ must refer only to shared variables in $\sigma$.  The value of $x$ is updated to the fully evaluated value, $\evalse$.  

\OMIT{
The second case includes the special read label $\Readxv$, which is generated at the process level by a load of $x$ \refrule{locals}.  
This transition is possible only when 
when $\sigma(x) = v$.  Note that transitions $\Readx{v'}$ generated at the process level where $\sigma(x) \neq v'$ have no possible transitions and
hence are eliminated from the traces of the system.  A load of $x$ makes no change to the global state and thus becomes an internal step%
\footnote{
For clarity and debugging purposes in the prototype tool we maintain register and variable names where appropriate in the promoted trace.
}.
}

\refrule{globals-guard} states that a guard transition $\guarde$ is possible exactly when $e$ evaluates to true in
the global state.  If it does not, no transition is possible; this is how incorrect branches are eliminated from the
traces.  If a guard does not evaluate to $true$, execution stops in the sense that no
transition is possible.  This corresponds to a false guard, i.e., $\Magic$ \cite{Morgan:94,Back:98}, and such
behaviours do not terminate and are ignored for the purposes of determining behaviour of a real system.
Interestingly, this straightforward concept from standard refinement theory allows us to handle speculative execution
straightforwardly.  In existing approaches, the semantics is complicated by needing to restart reads if speculation
proceeds down the wrong path.  Treating branch points as guards works because speculation should have no effect if the
wrong branch was chosen.  

\label{load-speculation}
\newcommand{\Mark}[1]{\underline{#1}}

To understand how this approach to speculative execution works, consider the following deri\-vation.  Assume that (a) loads may be reordered before guards if they reference independent variables, and (b) loads may be reordered if they reference different variables.  Recall that we omit trailing $\Skip$ commands to save space.
\begin{derivation}
	\step{
		r_1 \asgn x \cbef (\If r_1 = 0 \Then \Mark{r_2 \asgn y})
	}
	\trans{=}{Definition of $\If$ \refeqn{defn-if}}
	\step{
		r_1 \asgn x \cbef ((\guard{r_1 = 0} \cbef \Mark{r_2 \asgn y}) \choice \guard{r_1 \neq 0})
	}
	\trans{\refsto}{Resolve to the first branch, since $(\cmdc \choice \cmdd) \refsto \cmdc$ \OMIT{by \refeqn{defn-reft}}}
	\step{
		r_1 \asgn x \cbef \guard{r_1 = 0} \cbef \Mark{r_2 \asgn y} 
	}
	\trans{\refsto}{From \reflaw{swap-order} and assumption (a)} 
	\step{
		r_1 \asgn x \cbef \Mark{r_2 \asgn y} \bef \guard{r_1 = 0} 
	}
	\trans{\refsto}{From \reflaw{swap-order} and assumption (b)}
	\step{
		\Mark{r_2 \asgn y} \bef r_1 \asgn x \cbef \guard{r_1 = 0} 
	}
\end{derivation}
This shows that the inner load (underlined) may be reordered before the branch point, and subsequently before an earlier load.
Note that this behaviour results in a terminating trace only if $r_1 = 0$ holds when the guard is evaluated, and otherwise becomes $\Magic$ (speculation down
an incorrect path).
On ARM processors, placing a control fence ($\cfence$) instruction inside the branch, before the inner load prevents this reordering (
see  \refsect{reordering-forwarding}).

\subsubsection{Non-multi-copy atomic storage subsystem.}
\label{storage-subsystem-semantics-armv8}
Some versions of ARM and POWER allow processes to communicate values to each other without accessing the heap.  
That is, if process $\prp_1$ is storing $v$ to $x$, and process $\prp_2$ wants to
load $x$ into $r$, $\prp_2$ may preemptively load the value $v$ into $r$, before $\prp_1$'s store hits the global shared storage.  
Therefore different processes may have
different views of the value of a global variable, as exposed by litmus tests such as the \T{WRC} family \cite{HerdingCats}.  

Our approach to modelling this is based on that of the operational model of \cite{UnderstandingPOWER}.  However, that model maintains several
partial orders on operations reflecting the nondeterminism in the system, whereas we let the nondeterminism be represented by choices in the operational rules.
This means we maintain a simpler data structure, a single global list of writes.
The shared state from the perspective of a given process is a particular view of this list. There is no single definitive shared state.  In addition, viewing a value in the list causes the list to be updated 
and this affects later views.
To obtain the value of a variable this list is searched starting with the most recent write first.  A process $\prp_1$ that has already seen the latter of two updates to a variable $x$
may not subsequently then see the earlier update.  
Hence the list 
keeps track of which processes have seen which stores.  Furthermore, accesses of the storage subsystem are influenced by fences.

A write $w$ has the syntactic form $\srqnxvns$, where $x$ is a global variable being updated to value $v$,
$\pidn$ is the process id of the process from which the store originated, and $\pidns$ is the set of process ids that have ``seen''
the write. For such a $w$, we let $w.var=x$, $w.thread=\pidn$ and $w.seen=\pidns$.
For a write $\srqnxvns$ it is always the case that $\pidn \in \pidns$. 
The storage $\wseq$ is a list of writes, initially populated with writes for the initial values of global variables, which
all processes have ``seen''.

\begin{figure}[t]
\ruledef{120mm}
{storage-load}{
	\Rule{
		p \ttra{\pidn:\Readxv} p'
		\also
		\all w \in \ran(\wseq_1) @ x = w.var \imp \pidn \notin w.seen 
	}{
		\storage{\wseq_1 \cat \srqmxv{\pidns} \cat \wseq_2}{p}
		\ttra{\pidn:\Readxv}
		\storage{\wseq_1 \cat \srqmxv{\pidns \union \{\pidn\}} \cat \wseq_2}{p'}
	}
}

\vskip 3mm

\ruledef{120mm}
{storage-store}{
	\Rule{
		p \ttra{\pidn: x \asgnlbl v} p'
		\\
		\all w \in \ran(\wseq_1) @ 
			\pidn \neq w.thread 
			\land 
			(x = w.var \imp \pidn \notin w.seen)
	}{
		\storage{\wseq_1 \cat \wseq_2}{p}
		\ttra{\pidn: x \asgnlbl v}
		\storage{\wseq_1 \cat \srqnxvn \cat \wseq_2}{p'}
	}
}

\vskip 3mm

\ruledef{120mm}
{fence}{
	\Rule{
		p \ttra{\pidn:\fencelbl} p'
	}{
		\storage{\wseq}{p}
		\ttra{\pidn:\fencelbl}
		\storage{\flush_\pidn(\wseq)}{p'}
	}
}

where
\begin{equation*}
	\flush_\pidn(\eseq) = \eseq
	\qquad
	\flush_\pidn(w \cat \wseq) = 
		\left\{
		\begin{array}{ll}
			\Update{w}{seen}{\PIDSet} \cat \flush_\pidn(\wseq) 
			&
			if~ \pidn \in w.seen
			\\
			w \cat \flush_\pidn(\wseq) 
			& 
			otherwise
		\end{array}
		\right.
\end{equation*}
\caption{Rules for the non-multi-copy atomic subsystem of ARM and POWER}
\label{storage-rules}
\end{figure}

We give two specialised rules (for a load and store) in \reffig{storage-rules}.%
\footnote{
To handle the general 
case of an assignment $x \asgn e$, where $e$ may contain more than one shared variable, the 
antecedents of the rules are combined, retrieving the value of each variable referenced in $e$ individually and
accumulating the changes to $\wseq$.
}
\refrule{storage-load} states that a previous
write to $x$ may be seen by process $\pidn$ if there are no more recent writes to $x$ that it has already seen.
Its id is added to the set of processes that have seen that write.
\refrule{storage-store} states that
a write to $x$ may be added to the system by process $\pidn$, \emph{appearing earlier than existing writes in the system}, 
if the following two conditions hold for each of those existing writes $w$:
they are not by $\pidn$
($\pidn \neq w.thread$, local coherence), 
and
$x = w.var \imp \pidn \notin w.seen$, i.e., writes to the same variable are seen in a consistent order 
(although not all writes need be seen).

A $\fence$ action by process $\pidn$ `flushes' all previous writes by and seen by $\pidn$. 
The $\flush$ function modifies $\wseq$ so that all processes can see all writes by $\pidn$, effectively overwriting earlier writes.
This is achieved by updating the write so that all processes have seen it, written as
$\Update{w}{seen}{\PIDSet}$.

\subsection{Reordering and forwarding for ARM and POWER}
\label{reordering-forwarding}

\begin{figure}[t]
\begin{minipage}{0.45\linewidth}
\begin{eqnarray}
	\aca &\nro& \fence 
	\label{eq:a<f}
	\\
	\fence &\nro& \aca
	\label{eq:f<a}
	\\
	\guardb &\nro& \cfence
	\label{eq:g<cf}
	\\
	\cfence &\nro& r \asgn e
	\label{eq:cf<l}
	\\
	\guard{b_1} &\ro& \guard{b_2}
	\label{eq:g<g}
	\\
	\guardb &\nro& \gvar \asgn e
	\label{eq:g<s}
	\\
	\guardb &\ro& r \asgn e
		\ltiff{$r \nfi b$}
	\label{eq:g<r}
	\\
	x \asgn e &\ro& \guardb
		\ltiff{$x \nfi b$}
	\label{eq:u<g}
	\\
	\aca &\ro& \acb 
		~~
		\mbox{in all other cases}
		\notag
\end{eqnarray}
\end{minipage}
\begin{minipage}{0.53\linewidth}
\begin{eqnarray}
	x \asgn e &\ro& y \asgn f 
		\mbox{~iff}
	\label{eq:u<u}
		\\ &&
		\hspace{6pt}
		\mbox{$x \neq y$, $x \nfi f$, $y \nfi e$, and}
		\notag
		\\&&
		\hspace{6pt}
		\mbox{$e,f$ are load-distinct}
		\notag
	\\
	&& \notag \\
	\fwd{y \asgnlbl f}{x \asgn e} &=& x \asgn \Repl{e}{y}{f}
		\mbox{~if}
	\label{eq:fwd-u}
		\\ &&
		\hspace{6pt}
		\mbox{$e$ has no shared variables}
		\notag
	\\
	\fwd{y \asgnlbl f}{\guarde} &=& \guard{\Repl{e}{y}{f}}
		\mbox{~if}
	\label{eq:fwd-g}
		\\ &&
		\hspace{6pt}
		\mbox{$e$ has no shared variables}
		\notag
	\\
	\notag
	\fwd{\aca}{ \acb} &=& \acb
	~~
		\mbox{otherwise}
\end{eqnarray}
\end{minipage}
\caption{
Reordering and forwarding following ARM assembler semantics.
Let $x,y$ denote any variable, $r$ a local variable, and $\gvar$ a global variable.
}
\label{higher-level-defs}
\end{figure}

Our general semantics is instantiated for ARM and POWER processors in \reffig{higher-level-defs} which provides particular definitions for the reordering relation and forwarding that are generalised from the orderings on stores and loads in these processors.%
\footnote{We have excluded address shifting, which creates \emph{address dependencies} \cite{HerdingCats}, as 
this does not affect the majority of high-level algorithms in which we are interested. 
However, address dependencies are accounted for in our tool as discussed in \refappendix{address-shifting}.}

Fences prevent all reorderings \refeqns{a<f}{f<a}. 
Control fences prevent speculative loads when placed between a guard and a load \refeqns{g<cf}{cf<l}.  
Guards may be reordered with other guards \refeqn{g<g}, but stores to shared variables may not come before a guard evaluation \refeqn{g<s}.
This prevents speculative execution from modifying the global state, in the event that the speculation was down the wrong branch.
An update of a local variable may be reordered before a guard provided it does not affect the guard expression \refeqn{g<r}.
Guards may be reordered before updates if those updates do not affect the guard expression \refeqn{u<g}.

Assignments may be reordered as shown in \refeqn{u<u} and discussed in \refsect{overview-reordering}.
Forwarding is defined straightforwardly so that an earlier update modifies the expression of a later update or guard \refeqns{fwd-u}{fwd-g}, provided 
it references
no shared variables.

\section{Model checking concurrent data structures}
\label{tool}

Our semantics has been encoded in the Maude rewriting system \cite{Maude}. 
We have used the resulting prototype tool
to validate the semantics against litmus tests which have been used in other work on ARM (348 tests) \cite{ModellingARMv8} and POWER
(758 tests) \cite{UnderstandingPOWER}.  As that research was developed through testing on hardware and in consultation with the processor
vendors themselves we consider compliance with those litmus tests to be sufficient validation.  
With two exceptions, as discussed in \refsect{related-work},
our semantics agrees with
those results.  

We have employed Maude as a model checker to verify that a (test-and-set) lock provides mutual exclusion on ARM and POWER, and that
a lock-free stack algorithm,
and a deque (double-ended queue) algorithm, satisfy their
abstract specifications on ARM and POWER.  We describe the verification of the deque below, in which we found a bug in the published algorithm.

\subsection{Chase-Lev deque}
\label{chase-lev}

L\^{e} et. al \cite{LeWorkStealingPPoPP13} present a version of the  
Chase-Lev deque \cite{ChaseLev05} adapted for ARM and POWER.
The deque is implemented as an array, where elements may be \emph{put} on or \emph{taken} from the tail,
and additionally, processes may \emph{steal} an element from the head of the deque.
The \code{put} and \code{take} operations
may be executed by a single process only, hence there is no interference between these two operations
(although instruction reordering could cause consecutive
invocations to overlap).  The \code{steal} operation can be executed by multiple processes concurrently.

The code we tested is given in \reffig{chaselev-code} where $L$ is the maximum size of the deque which is implemented as a cyclic array, with
all elements initialised to some irrelevant value.
The original code includes handling array resizing, but here we focus on the insert/delete logic.  For brevity we omit trailing $\Skip$s.  We have
used a local variable $return$ to model the return value, and correspondingly
have refactored the algorithm to eliminate returns from within a branch.
A \code{\CAS{x}{ r}{ e}} (compare-and-swap) instruction atomically compares the value of global \code{x} with the value \code{r} and if the
same updates \code{x} to \code{e}.  We model a conditional statement with a \code{CAS} as follows.
\begin{equation}
\label{eq:tso-cas}
	\If \CAS{x}{ r}{ e} \Then c_1 \Else c_2
	\sdef
	(\atomic{\guard{x = r} \atomicsep x \asgn e} \cbef c_1 )
	\choice 
	(\guard{x \neq r} \cbef c_2 )
\end{equation}

\OMIT{
\begin{figure}[t]
\begin{lstlisting}
Initially head $\mapsto$ 0, tail $\mapsto$ 0, tasks $\mapsto$ $\seqT{\_, \_, \ldots}$

put(v) $\sdef$
  lcl t $\mapsto$ _ @
  t $\asgn$ tail ;
  tasks[t mod L] $\asgn$ v ;
  fence ;
  tail $\asgn$ t + 1
\end{lstlisting}

\vskip -3mm
\begin{lstlisting}[multicols=2]
 take $\sdef$
	lcl h $\mapsto$ _ , t $\mapsto$ _ , return $\mapsto$ _ @
	t $\asgn$ tail - 1 ;
	tail $\asgn$ t ;
	fence  ;
	h $\asgn$ head ;
	if h \leq t then ;
	  return $\asgn$ tasks[t mod L] ;
	  if h = t then
	  	if !CAS(head, h, h + 1) then
	  		return $\asgn$ empty  ;
	  	tail $\asgn$ t + 1  ;
	else
	  return $\asgn$ empty  ;
	  tail $\asgn$ t + 1

 steal $\sdef$
	lcl h $\mapsto$ _ , t $\mapsto$ _ , task $\mapsto$ _ , return $\mapsto$ _  @
	h $\asgn$ head ;
	fence  ;
	t $\asgn$ tail ;
	cfence ; // unnecessary 
	if h < t then ;
	  task $\asgn$ tasks[h mod L] ;
	  cfence ; // incorrectly placed
	  if CAS(head, h, h+1) then
	  	  return $\asgn$ task ;
	  else
	  	  return $\asgn$ fail ;
	else
	  return $\asgn$ empty
\end{lstlisting}
\caption{A version of L\^{e} et. al's work-stealing deque algorithm for ARM}
\label{chaselev-code}
\end{figure}
}

\begin{figure}[t]

\hspace{-5mm}
\parbox{0.87\textwidth}{
\[
\noindent
Initial~state: \{head \mapsto 0, tail \mapsto 0, tasks \mapsto \seqT{\_, \_, \ldots}\}
\\~\\
put(v) \sdef \\
  \quad \localsWord~ t \mapsto \_ @ \\
  \quad t \asgn tail \cbef \\
  \quad tasks[t \mod L] \asgn v \cbef \\
  \quad \fence \cbef \\
  \quad tail \asgn t + 1
\]
}
\newcommand{\IFWord}{\mathop{\mathbf{if}}}
\newcommand{\THENWord}{\mathop{\mathbf{then}}}
\newcommand{\ELSEWord}{\mathop{\mathbf{else}}}
\vskip -5mm
\hspace{-5mm}
\parbox{0.47\textwidth}{
\[
 take \sdef\\
	\quad \localsWord~ h \mapsto \_ , t \mapsto \_ , return \mapsto \_ @ \\
	\quad t \asgn tail - 1 \scomp \\
	\quad tail \asgn t \scomp \\
	\quad \fence  \scomp \\
	\quad h \asgn head \scomp \\
	\quad \IFWord h \leq t \THENWord \\
	  \quad \quad return \asgn tasks[t \mod L] \scomp \\
	  \quad \quad \IFWord h = t \THENWord \\
	  	\quad \quad \quad \IFWord \neg \CAS{head}{ h}{ h + 1} \THENWord \\
	  		\quad \quad \quad \quad return \asgn empty  \\
	  	\quad \quad \quad tail \asgn t + 1  \\
	\quad \ELSEWord \\
	  \quad \quad return \asgn empty  \scomp \\
	  \quad \quad tail \asgn t + 1 \\
\]
}
\hspace{-5mm}
\parbox{0.53\textwidth}{
\[
 steal \sdef \\
	\quad \localsWord~ h \mapsto \_ , t \mapsto \_ , return \mapsto \_  @ \\
	\quad h \asgn head \scomp \\
	\quad \fence  \scomp \\
	\quad t \asgn tail \scomp \\
	\quad \cfence \scomp \color{dkgreen}{//~unnecessary}  \\
	\quad \IFWord h < t \THENWord \\
	  \quad \quad return \asgn tasks[h \mod L] \scomp \\
	  \quad \quad \cfence \scomp \color{dkgreen}{//~incorrectly~placed} \\
	  \quad \quad \IFWord \neg \CAS{head}{ h}{ h+1} \THENWord \\
	  	  \quad \quad \quad return \asgn fail \\
	\quad \ELSEWord \\
	  \quad \quad return \asgn empty
	\\~\\
	~\\
\]
}

\caption{A version of L\^{e} et. al's work-stealing deque algorithm for ARM \cite{LeWorkStealingPPoPP13}}
\label{chaselev-code}
\end{figure}

The \code{put} operation straightforwardly adds an element to the end of the deque, incrementing the \code{tail} index.  It includes a full fence
so that the tail pointer is not incremented before the element is placed in the array.  The \code{take} operation uses a \code{CAS} operation
to atomically increment the head index.  Interference can occur if there is a concurrent \code{steal} operation in progress, which also uses
\code{CAS} to increment \code{head} to remove an element from the head of the deque.  The \code{take} and \code{steal} operation return empty if
they observe an empty deque.  In addition the \code{steal} operation may return the special value \code{fail} if interference on \code{head} occurs.
Complexity arises if the deque has one element and there are concurrent processes trying to both \code{take} and \code{steal} that element at the
same time.  

Operations \code{take} and \code{steal} use a \code{\fence} operation to ensure they have consistent readings for the head and tail indexes, and
later use \code{CAS} to atomically update the head pointer (only if necessary, in the case of \code{take}).  Additionally, the \code{steal}
operation contains two $\cfence$ barriers (\T{ctrl\_isync} in ARM).  


\subsubsection{Verification.}
\label{chaselev-verification}

\def\absput{put}
\def\abstake{take}
\def\abssteal{steal}

We use an abstract model of the deque and its operations to specify the allowed final values of the deque and return values.
The function \code{last(q)} returns the last element in \code{q} and \code{front(q)} returns \code{q} excluding its last element.
\begin{eqnarray*}
	\absput(v) &\sdef&
		q \asgn q \cat \seqT{v}
	\\
	\abstake &\sdef&
		\localsWord\ return \asgn none @ 
			\\ && \quad
		    \atomic{ \guard{q = \eseq} \atomicsep return \asgn empty }  \choice
			\\ && \quad
		    \atomic{ \guard{q \neq \eseq} \atomicsep return \asgn last(q) \atomicsep q \asgn front(q) }
	\\
	\abssteal &\sdef&
		\localsWord\ return \asgn none
			\\ && \quad
		    \atomic{ \guard{q = \eseq} \atomicsep return \asgn empty } \choice
			\\ && \quad
		    \atomic{ \guard{q \neq \eseq} \atomicsep return \asgn head(q) \atomicsep q \asgn tail(q) }
\end{eqnarray*}
The abstract specification for \code{steal} does not attempt to detect interference and return \code{fail}.  As such we exclude these behaviours of the concrete code from the analysis.

We model-checked combinations of one to three processes operating in parallel, each executing one or two operations in sequence.  The final states of the
abstract and concrete code were compared via a simulation relation.
This exposed a bug in the code which may occur when a \code{put} and \code{steal} operation execute in parallel on an empty deque.  
The load 
$return \asgn tasks[h \mod L]$ can be speculatively executed before the guard $h < t$ is evaluated, and hence also before the load of \code{tail}.  Thus the steal
process may load \code{head}, load an irrelevant \code{return} value, at which point a \code{put} operation may complete, storing a value and incrementing \code{tail}.  The
\code{steal} operation resumes, loading the new value for \code{tail} and observing a non-empty deque, succeeding with its \code{CAS} and returning the irrelevant value, which was loaded before
the \code{put} operation had begun.

Swapping the order of the second \code{\cfence} with the load of $tasks[h \mod L]$ eliminates this bug, and our analysis did not reveal any other problems.  In addition,
eliminating the first \code{\cfence} does not change the possible outcomes.  

%

\section{Related work}
\label{related-work}

This work makes use of an extensive suite of tests elucidating the behaviour of weak memory models in ARM and POWER via both operational and axiomatic semantics
\cite{HerdingCats,UnderstandingPOWER,AxiomaticPower,ModellingARMv8}. Those semantics were developed and validated through testing on real hardware and in consultation with
processor vendors themselves.  Our model is validated against their results, in the form of the results of litmus tests.  

Excluding two tests involving ``shadow registers'', which appear to be pro\-cessor-specific facilities which are not intended
to conform to sequential semantics (they do not correspond to higher-level code), 
all of the 348 ARM litmus tests run on our model agreed with the
results in \cite{ModellingARMv8},
and all of the 758 POWER litmus tests run on our model agreed with the
results in \cite{UnderstandingPOWER}, which the exception of litmus test
\T{PPO015},
which we give below, translated into our formal language.%
\footnote{
We simplified some of the syntax for clarity, in particular introducing a higher-level $\If$ statement to model a jump
command and implicit register (referenced by the compare (\T{CMP}) and branch-not-equal (\T{BNE}) instructions).
We have also combined some commands, retaining dependencies, in a way that is not possible in the assembler language.
The \T{xor} operator is exclusive-or; its use here artificially creates a \emph{data dependency} \cite{HerdingCats} between the updates to $r_0$ and $z$.
}
\OMIT{
\T{PPO015} is shown below. 
We have renamed some variables and simplified some of the syntax for clarity, in particular introducing a higher-level $\If$ statement to remove a jump
command and implicit register (referenced by the compare (\code{CMP}) and branch-not-equal (\code{BNE}) instructions).
The key point is that the expression of the $\If$ statement references register \code{r_3} which holds the loaded value of $z$.  Hence the branch cannot be
evaluated (non-speculatively) until the value for $z$ is known.
\begin{equation}
\begin{split}
\label{eq:1ppo015}
	~&
 \storex{1}
 \cbef
 \fence
 \cbef
 \storey{1}
 \quad
 \pl
 \\
 	~&
 \load{r0}{y}  
 \cbef
  \mathttbf{eor}~\T{r1, r0, r0}
 \cbef
  \mathttbf{add}~\T{r1, r1, 1}
 \cbef
  \store{z}{r1}
 \cbef
  \store{z}{2}
 \cbef
  \load{r3}{z}
 \cbef
  \\ 
  	~&
  \qquad \qquad
  (\If \T{r3 = r3} \Then \Skip \Else \Skip)
 \cbef
  \cfence
 \cbef
  \load{r4}{x}
\end{split}
\end{equation}
}
\begin{equation}
\begin{split}
\label{eq:1ppo015}
	~&
 x \asgn 1
 \cbef
 \fence
 \cbef
 y \asgn 1
 \quad
 \pl
 \\
 	~&
 r_0 \asgn y
 \cbef
 z \asgn (r_0~{\tt xor}~r_0) + 1
 \cbef
 z \asgn 2
 \cbef
 r_3 \asgn z
 \cbef
  \\ 
  	~&
  \qquad \qquad
  (\If r_3 = r_3 \Then \Skip \Else \Skip)
 \scomp
  \cfence
 \cbef
  r_4 \asgn x
\end{split}
\end{equation}
The tested condition is 
$z=2 \land r_0=1 \land r_4=0$, which asks whether it is possible to load $x$ (the last statement of process 2) before loading $y$ (the first statement of
process 2). At a first glance the control fence prevents the load of $x$ happening before the branch.  However, as indicated by litmus tests such as
\T{MP+dmb.sy+fri-rfi-ctrisb}, 
\cite[Sect 3,\emph{Out of order execution}]{ModellingARMv8}, 
under some circumstances the branch condition can be
evaluated early, as discussed in the speculative execution example.  
We expand on this below by manipulating the second process, taking the case where the success branch of the $\If$ statement is chosen.
To aid clarity we underline the instruction that is the target of the (next) refinement step.
\begin{derivation*}
\OMIT{
	\step{
 r_0 \asgn y
 \cbef
 z \asgn (r_0~{\tt xor}~r_0) + 1
 \cbef
 z \asgn 2
 \cbef
 r_3 \asgn z
 \cbef
  (\underline{\guard{r_3 = r_3}} \choice \guard{r_3 \neq r_3}) 
 \scomp
  \cfence
 \cbef
  r_4 \asgn x
	}

	\trans{\refsto}{Choose first branch}
}

	\step{
 r_0 \asgn y
 \cbef
 z \asgn (r_0~{\tt xor}~r_0) + 1
 \cbef
 z \asgn 2
 \cbef
 \underline{r_3 \asgn z}
 \cbef
  \guard{r_3 = r_3} 
 \cbef
  \cfence
 \cbef
  r_4 \asgn x
	}

	\trans{\refsto}{Promote load with forwarding (from $z \asgn 2$), from \reflaws{keep-order}{swap-order}}

	\step{
 r_3 \asgn 2
 \bef
 r_0 \asgn y
 \cbef
 z \asgn (r_0~{\tt xor}~r_0) + 1
 \cbef
 z \asgn 2
 \cbef
  \underline{\guard{r_3 = r_3} }
 \cbef
  \cfence
 \cbef
  r_4 \asgn x
	}

	\trans{\refsto}{Promote guard by \reflaws{keep-order}{swap-order} (from \refeqn{u<g})}

	\step{
 r_3 \asgn 2
 \bef
  \guard{r_3 = r_3} 
 \bef
 r_0 \asgn y
 \cbef
 z \asgn (r_0~{\tt xor}~r_0) + 1
 \cbef
 z \asgn 2
 \cbef
  \underline{\cfence}
 \cbef
  r_4 \asgn x
	}

	\trans{\refsto}{Promote control fence by \reflaws{keep-order}{swap-order} (\refeqn{g<cf} does not now apply)}

	\step{
 r_3 \asgn 2
 \bef
  \guard{r_3 = r_3} 
 \bef
  \cfence
 \bef
 r_0 \asgn y
 \cbef
 z \asgn (r_0~{\tt xor}~r_0) + 1
 \cbef
 z \asgn 2
 \cbef
  \underline{r_4 \asgn x}
	}

	\trans{\refsto}{Promote load by \reflaws{keep-order}{swap-order} }

	\step{
 r_3 \asgn 2
 \bef
  \guard{r_3 = r_3} 
 \bef
  \cfence
 \bef
  r_4 \asgn x
 \bef
 r_0 \asgn y
 \cbef
 z \asgn (r_0~{\tt xor}~r_0) + 1
 \cbef
 z \asgn 2
	}

\end{derivation*}
The load $r_4 \asgn x$ has been reordered before the load $r_0 \asgn y$, and hence when interleaved with the first process from \refeqn{1ppo015}
it is straightforward that the condition may be satisfied.

\OMIT{
In our model, the instruction $\load{r3}{z}$ can be reordered before $\store{z}{2}$, which forwards the value 2 to \code{r3}.  Now
$\reg{r3}{2}$ can be executed (reordered before all earlier instructions), and subsequently so can the guard instruction $\guard{r3 = r3}$. 
This allows the control fence to be executed since there is no preceding branch point, 
and finally $\load{r4}{x}$, since it is not dependent on any earlier register or shared variable and is no longer blocked by the control fence.  
As such the value loaded into $r4$ may be the initial value for $x$.  The process 1 then executes to completion, after which process 2 continues by loading the value 1 for $y$.
}

In the Flowing/POP model of \cite{ModellingARMv8}, 
this behaviour is forbidden because there is a data dependency from the load of $y$ into \code{r_0} to \code{r_3}, via $z$.  This
appears to be because of the consecutive stores to $z$, one of which depends on \code{r_0}.  
In the testing of real processors reported in \cite{ModellingARMv8}, the behaviour that we allow was never observed, but is allowed by the model
in \cite{HerdingCats}.  
\OMIT{
This may be because that behaviour
is intended to be allowed by the specification of the architecture, but is not realised on any processor.  However, it may also be intended to be
disallowed, in which case our model is wrong.

If the behaviour of litmus test \T{PPO015} \refeqn{1ppo015} is intended to be forbidden, that requires forwarding values only when earlier stored values have been fully
determined (i.e., when the value for \T{r1} is known).  That cannot be determined with the simple pair-wise comparison that our framework uses.
It would also appear to limit the amount of reordering possible, for no change in sequential semantics (it is still valid to forward the known
value 2 for $z$ to later loads of $z$).  
}
As such we deem this discrepancy to be a minor issue in Flowing/POP (preservation of transitive
dependencies) rather than a fault in our model.

Our model of the storage subsystem is similar to that of the operational models of
\cite{UnderstandingPOWER,ModellingARMv8}.  However our thread model is quite different, being defined in terms of relationships between
actions.  The key difference is how we handle branching and the effects of speculative execution.  The earlier models are complicated in
the sense that they are closer to the real execution of instructions on a processor, involving restarting reads if an earlier read
invalidates the choice taken at a branch point.  


The axiomatic models, as exemplified by \cite{HerdingCats}, define relationships between instructions in a whole-system way, including relationships between
instructions in concurrent processes.  This gives a global view of how an architecture's reordering rules (and storage system) interact to reorder instructions in a
system.  Such global orderings are not immediately obvious from our pair-wise orderings on instructions.  On the other hand, those globals orderings become quite
complex and obscure some details, and it is unclear how to extract some of the generic principles such as \refeqn{reordering-principle}.

\section{Conclusion}\label{conclusion}

We have utilised earlier work to devise a wide-spectrum language and semantics for weak memory models which is relatively straightforward to define and
extend, and which lends itself to verifying low-level code against abstract specifications.  While abstracting away from the details of the architecture,
we believe it provides a complementary insight into why some reorderings are allowed, requiring a pair-wise relationship between instructions rather than
one that is system-wide.  

A model-checking approach based on our semantics exposed a bug in an algorithm in \cite{LeWorkStealingPPoPP13} in relation to the placement of a
control fence.  The original paper includes a hand-written proof of the correctness of the algorithm based on the axiomatic model of
\cite{AxiomaticPower}.  The possible traces of the code were enumerated and validated against a set of conditions on adding and removing
elements from the deque (rather than with respect to an abstract specification of the deque).  The conditions being checked are non-trivial
to express using final state analysis only.  An advantage of having a
semantics that can apply straightforwardly to abstract specifications, rather than a proof technique that analyses behaviours of the
concrete code only, is that we may reason at a more abstract level.


We have described the ordering condition as syntactic constraints on atomic actions, which fits with the low level decisions of hardware
processors such as ARM and POWER.  However
our main reordering principle \refeqn{reordering-principle} is
based on semantic concerns, and as such may be applicable as a basis for understanding the
interplay of software memory models, compiler optimisations and hardware memory models \cite{PromisingSemantics}.

The wide-spectrum language has as its basic instruction an assignment, which is sufficient for specifying
many concurrent programs.  However we hope to extend the language
to encompass more general constructs
such as the specification command \cite{SpecStmt} and support rely-guarantee reasoning \cite{Jones-RG1,Jones-RG2,SRA16,SemanticsSRA}.\\

\noindent{\bf Acknowledgements} We thank Kirsten Winter for feedback on this work, and the support of Australian Research Council Discovery Grant DP160102457.

\OMIT{
\newpage
\section{Related work; updated.  Perhaps ``Validation on hardware''}

Previous formalisations of the semantics of ARM include \cite{HerdingCats,ModellingARMv8}.  The former was validated against a total of 9790 litmus tests.  
We exclude 3626 as they use address shifting, which we have not fully accounted for.  
We also exclude a test involving shadow registers, which we do not handle.  
A further 29 (37 - 8 addr) could not be automatically parsed and have been excluded from the analysis.  
We exclude a further 136 (142 - 6 addr)
as our encoding in Maude was unable to finish due to memory limitations.  

This leaves 5998 tests.  
Of the remaining tests, our results disagree with that of \cite{HerdingCats}
on 41 tests; however, in these 41 cases our results accord with those observed on hardware (40 where the herding cats model allows the behaviour and we do not, and one
where it is allowed on hardware but forbidden by their model; our model also allows it).  
This leaves 6 of the 5998 where our results disagree with both the Herding Cats model and the hardware results%
\footnote{
\raggedright
{\tt 
DETOUR0692,
DETOUR0706,
DETOUR1157,
DETOUR1167,
DETOUR1171,
S+dmb+data-wsi
}}%
.  In each case, the model forbids the behaviour and it has
not been observed on hardware, but out model allows it.  (There are no instances of the case where the hardware allows the behaviour but we forbid it, except where that
agrees with the herding cats result.)

As reported in \cite{HerdingCats}, there are 2086 (2088 - 2 unknown) tests (of 9790) where their results disagree with that of hardware observations.  1532 are
allow/forbid and 554 are forbid/allow cases.  The majority of these cases were validated with ARM representatives who confirmed the results, not withstanding the
complication of some processors containing faults\footnote{
\raggedright
\url{http://infocenter.arm.com/help/topic/com.arm.doc.uan0004a/UAN0004A_a9_read_read.pdf}
}%
.

Note that of the 3626 Litmus tests involving address shifting, 50 failed (and 21 of those included discrepancies between the herding cats model and hardware results).

We also compare our results to those of the more recent \cite{ModellingARMv8}.  That includes 362 litmus tests, 4 of which we could not parse, and one involving shadow
registers.  Of the remaining 357, in all but one case our model disagreed with theirs (Flowing/POP): our model allows the tested condition, while theirs forbids it%
\footnote{{\tt PPO0015}}.
Interestingly, the herding cats model also allows it, and hence there is a discrepancy between those two models as well.  (The 357 tests include 84 involving address
shifting, all of which our model agrees with.)

\subsubsection{Future work.}
There are many similarities between POWER and ARM processors.  Our semantics is potentially applicable to POWER as well, therefore, and indeed we have so far showed
conformance with 758 tests reported in \cite{AxiomaticPower}, excluding the result for the same single discrepancy reported above for the Flowing/POP model
\cite{ModellingARMv8}.  POWER included a lightweight fence instruction, and included in the 757 are 395 litmus tests involving lightweight fences which we account for;
however we have more work to do to conform with the further tests reported for power in \cite{HerdingCats}.  We also intend to validate our semantics against TSO
\cite{x86-TSO}.
}

\bibliographystyle{plain}

\bibliography{biblio,colvinpubs}

\appendix

\newpage

\section{Lightweight fences}
\label{power}

POWER's \emph{lightweight fences} maintain order between loads, loads then stores, and stores, but not stores and subsequent loads
(loads can come before earlier stores).  If lightweight fences did not maintain
load-load order it would be straightforward to define their effect in terms of one instruction.  However to allow
reordering later loads with earlier stores but not earlier loads we model a $\lfence$ as two ``gates'',
one blocking loads and one stores.%
\OMIT{
It is possible to do it with one fence that marks any later load which `jumps' it so that load can be
reordered with earlier stores but \emph{not} earlier loads, however this seems less elegant and less 
amenable to algebraic analysis.
}

\renewcommand{\load}[2]{#1 \asgn #2}
\renewcommand{\store}[2]{#1 \asgn #2}
\renewcommand{\loadlbl}[2]{#1 \asgnsmall #2}

We define $\lfence \cbef c$ as
$
	\loadgate \cbef \storegate \cbef c
$
where
\begin{equation}
\begin{split}
\label{eq:lwfence-ro}
		\loadrx \nro \loadgate 
		\qquad
		\loadgate \nro \loadrx
		\\
		\storexv \nro \storegate 
		\qquad
		\storegate \nro \storexv
\end{split}
\end{equation}
Consider the code $c \cbef \loadgate \cbef \storegate \cbef d$ for arbitrary $c$ and $d$.  Without the intervening gates that constitute a lightweight fence the
instructions of $c$ and $d$ could be reordered according to the usual restrictions.  The $\storegate$ instruction however prevents stores in $d$
from interleaving with instructions in $c$ (but loads may be reordered past the $\storegate$ instruction).  Additionally, the $\loadgate$ may be
reordered past any stores in $c$ but not past any loads.  Hence between the two fence instructions can mix stores from $c$ and loads from $d$,
which may be reordered subject to the usual constraints.  The lightweight fence therefore maintains store/store, load/load, and load/store order.

A lightweight fence also has a global effect on the storage system, which we encode in the semantics of the
$\storegate$ instruction.
%
%
A lightweight fence by $\pidn$ marks any store in $\wseq$ seen by $\pidn$ with a tag $\lwfenced{\pidn}$. A store $w$ by process $\pidn$ may not be inserted in $\wseq$ before
a write with the $\lwfenced{\pidn}$ tag.  In addition, if another process $\pidm$ loads a value stored by $\pidn$, it \emph{sees}
not only that store but also all stores marked with $\lwfenced{\pidn}$. This transitive effect gives cumulativity of lightweight fences \cite{HerdingCats}. 
\newcommand{\lwflush}{lwflush}
%
%

As $\storegate$ is the latter to reach the storage system we give it the global effect; a $\loadgate$ instruction has no global effect.
\begin{equation}
	\Rule{
		p \ttra{\pidn:\storegatelbl} p'
	}{
		\storage{\wseq}{p}
		\ttra{\pidn:\storegatelbl}
		\storage{\lwflush_\pidn(\wseq)}{p'}
	}
\end{equation}
where
\begin{eqnarray*}
	\lwflush_\pidn(\eseq) &=& \eseq
	\\
	\lwflush_\pidn(\srqmxvns \cat \wseq) &=& 
	\left\{
		\begin{array}{ll}
		\srq{\pidm}{x}{v}{\lwfenced{\pidn} \cat \pidns} \cat \lwflush_\pidn(\wseq) 
		& if~ \pidn \in \pidns
	\\
		\srqmxvns \cat \lwflush_\pidn(\wseq) 
		& otherwise
	\end{array}
	\right.
\end{eqnarray*}
Adding the $\lwfenced{\pidn}$ tag to the list of process ids that have observed a write affects the allowed ordering of how writes are seen. 
The key point is that $\pidn$ now sees, and lightweight-fences, writes by $\pidm$ that $\pidm$ has lightweight-fenced.


\OMIT{
We define reordering on writes in the storage system as follows.
\begin{eqnarray*}
	\canReorder{\srqnxvn}{\srq{\pidm}{y}{v'}{\pidns}}
	&\iff&
	\pidm \neq \pidn
	\land
	\\&&
	(x = y \imp \pidn \notin \pidns )
	\land
	\lwfencedn \notin \pidns
\end{eqnarray*}
}
\OMIT{
where the final condition can be equivalently stated as
\begin{equation}
	\lnot(
	\pidm = \pidn
	\lor
	(x \neq y \land \pidn \in \pidns )
	\lor
	\lwfencedn \in \pidns
	)
\end{equation}
Note the difference now between lightweight fencing and seeing some other variable.  The fencing has a global effect on all orderings.
The condition  $\pidm \neq \pidn$ enforces the thread-local (re)ordering of writes to be maintained globally.
The condition $(x = y \imp \pidn \notin \pidns)$ enforces global coherence, so that writes to the same variable are seen in a consistent order 
(although not all writes need be seen).
The condition $\lwfencedn \notin \pidns$ maintains the order designated by lightweight fences, which may be across different variables.
}

The antecedent for
\refrule{storage-store}
needs to be updated to include $\lwfencedn \notin w.seen$ as a further constraint on where writes can be placed in the global order $\wseq$: a write may not
come before a write that the process has lightweight-fenced, even if that write is to a different variable.


\section{Address shifting} \label{address-shifting} 

In ARM (and POWER) the value loaded from (or stored to) an address may be shifted.
For the majority of high-level algorithms such details are hidden.  However address shifting is investigated at the hardware level because it can affect reordering --
so called ``address dependencies'' \cite{HerdingCats}.  The instruction \T{LDR~R1,\,[R2,\,X]} loads into \T{R1} the value at address \T{X} shifted by the amount in \T{R2}.  To
precisely model the semantics of address shifting requires a more concrete model than the one we propose, however, as determined by the litmus tests of
\cite{ModellingARMv8}, the effects of address dependencies can be investigated even when the shift amount is 0 (resulting in a load of the value at the address).
As such we define that an address shift of 0 on a variable $x$ gives $x$, and leave the effect of other shift amounts undefined.  

Address dependencies constrain the reorderings in the following ways: a branch
may not be reordered before a load or store with an (unresolved) address dependency; a store may not be reordered before an
instruction with an (unresolved) address dependency; and any instruction $\aca$ which shares a register or variable with $\acb$ where $\acb$ has
an unresolved address dependency may not be reordered with $\acb$.  We incorporate these conditions into the general rule for assignments.

A further consequence of address shifting is that a load $\load{r_2}{x}$ may be reordered before $\load{r_1}{\addrShift{x}{n}}$ even though this would 
violate load-distinctness.  However, to preserve coherence-per-location, the load into $r_2$ must not load a value of $x$ that was written before the value read
by the load into $r_1$.  This complex situation is handled in \cite{ModellingARMv8} by restarting load instructions if an earlier value is read into $r_2$.  We handle it more abstractly by
treating the load as speculation, where if an earlier value for $r_2$ is loaded then the effect of that speculation is thrown away.

\OMIT{
\begin{equation}
	\load{r_1}{\addrShift{x}{n}} \cbef \load{r_2}{x} \cbef c
	\leadsto
	\load{r_2}{x} \cbef \load{r_1}{\addrShift{x}{n}} \cbef \guard{r_1 = r_2} \cbef c
\end{equation}
}


We can give this extra semantics by adding an extra operational rule which applies only in those specific circumstances.
\begin{equation}
	\Rule{
		\prp \ttra{\loadlbl{r_2}{x}} \prp'
	}{
		\load{r_1}{\addrShift{x}{n}} \cbef \prp
		\ttra{\loadlbl{r_2}{x}}
		\load{r_1}{\addrShift{x}{n}} \cbef \guard{r_1 = r_2} \cbef \prp'
	}
\end{equation}
In practical terms it is possible the first load of $x$ (into $r_1$) is delayed while determining the offset value.
The later load is allowed to proceed, freeing up $\prp'$ to continue speculatively executing until the dependency is resolved.  The load
into $r_1$ then must still be issued, the result being checked against $r_2$.  This check must occur as to preserve coherency as the load into $r_2$ cannot
read a value earlier than that read into $r_1$.  Note that loads in $\prp'$ can now potentially be reordered to execute ahead of the load into $r_1$.

\OMIT{
Algebraically:
\[
	\load{r_1}{x} \cbef (\load{r_2}{x} \bef \prp)
	= 
	\load{r_2}{x} \bef 
	(\load{r_1}{x} \cbef (\guard{r_1 = r_2} \cbef \prp'))
\]
	
}

\end{document}